\documentclass[10pt]{article}

\usepackage{graphicx}
\usepackage{cite} 
 
\begin{document}


\title{CRNPRED: Highly Accurate Prediction of One-dimensional Protein Structures by Large-scale Critical Random Networks}
 

\author{Akira R Kinjo$^{1,2}$ and Ken Nishikawa$^{1,2}$\\
    $^1$Center for Information Biology and DNA Data Bank of Japan,\\
        National Institute of Genetics, Mishima, 411-8540, Japan\\
    $^2$Department of Genetics, The Graduate University for Advanced Studies (SOKENDAI), \\Mishima 411-8540, Japan
      }
      
\maketitle

\begin{abstract}
\textbf{Background:} 
One-dimensional protein structures such as secondary structures or contact 
numbers are useful for three-dimensional structure prediction and helpful for 
intuitive understanding of the sequence-structure relationship.
Accurate prediction methods will serve as a basis for these and other purposes.\\
\textbf{Results:} We implemented a program CRNPRED which predicts secondary 
structures, contact numbers and residue-wise contact orders. This program is 
based on a novel machine learning scheme called critical random 
networks. Unlike most conventional one-dimensional structure prediction 
methods which are based on local windows of an amino acid sequence, CRNPRED 
takes into account the whole sequence. CRNPRED achieves, on average per chain, 
$Q_3$ = 81\% for secondary structure prediction, and correlation coefficients 
of 0.75 and 0.61 for contact number and residue-wise contact order 
predictions, respectively.\\
\textbf{Conclusion:} CRNPRED will be a useful tool for computational as 
well as experimental biologists who need accurate one-dimensional protein 
structure predictions. 
\end{abstract}

\section*{Background}

One-dimensional (1D) structures of a protein are residue-wise 
quantities or symbols onto which some features of the native 
three-dimensional (3D) structure are projected.
1D structures are of interest for several reasons. For example, predicted 
secondary structures, a kind of 1D structures, are often used to limit the 
conformational space to be searched in 3D structure prediction. 
Furthermore, it has recently been shown that certain sets of the native 
(as opposed to predicted) 1D structures of 
a protein contain sufficient information to recover the native 3D 
structure~\cite{PortoETAL2004,KinjoANDNishikawa2005}. These 1D structures are
either the principal eigenvector of the contact map~\cite{PortoETAL2004} or a set of secondary structures (SS), contact numbers (CN) and residue-wise contact orders (RWCO)~\cite{KinjoANDNishikawa2005}.
Therefore, it is possible, at least in principle, to predict the native 3D 
structure by first predicting the 1D structures, and then by constructing 
the 3D structure from these 1D structures. 1D structures are not only useful 
for 3D structure predictions, but also helpful for intuitive understanding 
of the correspondence between the protein structure and its amino acid sequence
due to the residue-wise characteristics of 1D structures. Therefore, accurate 
prediction of 1D protein structures is of fundamental biological interest. 

Secondary structure prediction has a long history \cite{Rost2003}. 
Almost all the modern predictors are based on position-specific scoring 
matrices (PSSM) and some kind of machine learning techniques such as neural 
networks or support vector machines. Currently the best predictors achieve 
$Q_3$ of 77--79\% \cite{Jones1999,PollastriANDMcLysaght2005}. 
The study of contact number prediction also started long time 
ago \cite{NishikawaANDOoi1980, NishikawaANDOoi1986}, but further 
improvements were made only recently \cite{KinjoETAL2005, Yuan2005, KinjoANDNishikawa2005c}. These recent methods are based on the ideas developed in SS 
predictions (i.e., PSSM and machine learning), and achieve a correlation 
coefficient of 0.68--0.73.

Recently, we have developed a new method for accurately predicting SS, CN, 
and RWCO based on a novel machine learning scheme, 
critical random networks (CRN) ~\cite{KinjoANDNishikawa2005c}. 
In this paper, we briefly describe the formulation of the method, and recent 
improvements leading to even better predictions.
The computer program for SS, CN, and RWCO prediction named CRNPRED has been 
developed for the convenience of the general user, and a web interface and 
source code are made available online.
 
\section*{Implementation}

\subsection*{Definition of 1D structures}
\textit{Secondary structures (SS):}
Secondary structures were defined by the DSSP program \cite{DSSP}.
For three-state SS prediction, the simple encoding scheme (the so-called CK 
mapping) was employed \cite{CrooksANDBrenner2004}.
That is, $\alpha$ helices ($H$), $\beta$ strands ($E$), and other structures
(``coils'') defined by DSSP were encoded as $H$, $E$, and $C$, respectively.
Note that we do not use the CASP-style conversion scheme (the so-called EHL 
mapping) in which DSSP's $H$, $G$ ($3_{10}$ helix) and $I$ ($\pi$ helix) are encoded as $H$, and DSSP's $E$ and $B$ ($\beta$ bridge) as $E$.
We believe the CK mapping is more natural and useful for 3D structure 
predictions (e.g., geometrical restraints should be different between an 
$\alpha$ helix and a $3_{10}$ helix).
For SS prediction, we introduce feature variables $(y_i^H, y_i^E, y_i^C)$ 
to represent each type of secondary structures at the $i$-th residue position,
so that $H$ is represented as $(1,-1,-1)$, $E$ as $(-1,1,-1)$, and $C$ as 
$(-1,-1,1)$.

\textit{Contact numbers (CN):}
Let $C_{i,j}$ represent the contact map of a protein. Usually, the contact 
map is defined so that $C_{i,j} = 1$ if the $i$-th and $j$-th residues are in 
contact by some definition, or $C_{i,j} = 0$, otherwise. As in our 
previous study, we slightly modify the definition using a sigmoid function. 
That is, 
\begin{equation}
  C_{i,j} = 1/\{1+\exp[w(r_{i,j} - d)]\}
\end{equation}
where $r_{i,j}$ is the distance between $C_{\beta}$ ($C_{\alpha}$ 
for glycines) atoms of the $i$-th and $j$-th residues, $d = 12$\AA{} is a 
cutoff distance, and $w$ is a sharpness parameter of the sigmoid function 
which is set to 3 \cite{KinjoETAL2005,KinjoANDNishikawa2005}. The rather 
generous cutoff length of 12\AA{} was shown to optimize the prediction 
accuracy \cite{KinjoETAL2005}. The use of the sigmoid function enables us to 
use the contact numbers in molecular dynamics 
simulations \cite{KinjoANDNishikawa2005}.
Using the above definition of the contact map, the contact number of the
$i$-th residue of a protein is defined as
\begin{equation}
  n_i = \sum_{j:|i-j|>2}C_{i,j}. \label{eq:defcn}
\end{equation}
The feature variable $y_i$ for CN is defined as $y_i = n_i / \log L$ where 
$L$ is the sequence length of a target protein. The normalization 
factor $\log L$ is introduced because we have observed that the contact 
number averaged over a protein chain is roughly proportional to $\log L$,
and thus division by this value removes the size-dependence of predicted
contact numbers.

\textit{Residue-wise contact orders (RWCO):}
RWCO was first introduced in \cite{KinjoANDNishikawa2005}.
This quantity measures the extent to which a residue makes long-range contacts
in a native protein structure.
Using the same notation as contact numbers, 
the RWCO of the $i$-th residue in a protein structure is defined by 
\begin{equation}
  o_i = \sum_{j:|i-j|>2}|i-j|C_{i,j}. \label{eq:defrwco}
\end{equation}
The feature variable $y_i$ for RWCO is defined as $y_i = o_i / L$ where 
$L$ is the sequence length. Due to the similar reason as CN, the normalization
factor $L$ was introduced to remove the size-dependence of the predicted
RWCOs (the RWCO averaged over a protein chain is roughly proportional to the 
chain length).

\subsection*{Critical random networks}
Here we briefly describe the critical random network (CRN) method introduced 
in \cite{KinjoANDNishikawa2005c} which should be referred to for the details.
 Unlike most conventional methods for 1D structure prediction [except for 
some including the bidirectional recurrent neural networks \cite{BaldiETAL1999,PollastriANDMcLysaght2005,ChenANDChaudhari2006}], the CRN method 
takes the whole amino acid sequence into account. In the CRN method, 
an $N$-dimensional state vector $\mathbf{x}_i$  is assigned to the $i$-th 
residue of the target sequence (we use $N = 5000$ throughout this paper). 
Neighboring state vectors along the sequence 
are connected via a random $N\times N$ orthogonal matrix $W$. This matrix is 
also block-diagonal with the size of blocks ranging uniformly randomly 
between 2 and 50. The input to the CRN is the position-specific scoring matrix 
(PSSM), $U = (\mathbf{u}_1, \cdots, \mathbf{u}_L)$ 
of the target sequence obtained by PSI-BLAST~\cite{AltschulETAL1997} ($L$ is the sequence length of the target protein). 
We impose that the state vectors satisfy the following equation of state:
\begin{equation}
  \label{eq:eos}
  \mathbf{x}_i = \tanh[\beta W (\mathbf{x}_{i-1} + \mathbf{x}_{i+1}) + \alpha V\mathbf{u}_i]
\end{equation}
for $i = 1, \cdots , L$ where $V$ is an $N\times 21$ random matrix 
(the 21st component of $\mathbf{u}_i$ is always set to unity), and $\beta$ and $\alpha$ are scalar parameters. The fixed boundary condition is imposed ($\mathbf{x}_0 = \mathbf{x}_{L+1} = \mathbf{0}$). By setting $\beta = 0.5$, 
the system of state vectors is made to be near a critical point in a certain 
sense, and thus the range of site-site correlation is expected to be long 
when $\alpha$ is sufficiently small but finite~\cite{KinjoANDNishikawa2005c}. 
In this way, each state vector implicitly incorporates long-range correlations.
The 1D structure of the $i$-th residue is predicted as 
a linear projection of a local window of the PSSM and the state vector obtained by solving Eq. \ref{eq:eos}: 
\begin{equation}
  \label{eq:pred}
  y_i = \sum_{m=-M}^{M}\sum_{a=1}^{21}D_{m,a}u_{a,i+m} + \sum_{k=1}^{N}E_{k}x_{k,i}
\end{equation}
where $y_i$ is the predicted quantity, and $D_{m,a}$ and $E_k$ are the 
regression parameters. In the first summation, each PSSM column is extended to 
include the ``terminal'' residue. 
Since Eq. \ref{eq:pred} is a simple linear equation once the equation of 
state (Eq. \ref{eq:eos}) has been solved, learning the parameters $D_{m,a}$ and
 $E_{k}$ reduces to an ordinary linear regression problem.
For SS prediction, the triple $(y^{H}_i, y^{E}_i, y^{C}_i)$ is 
calculated simultaneously, and the SS class is predicted as 
$\mathrm{arg}\max_{s\in \{H, E, C\}}y^{s}_i$.  For the CN and RWCO prediction,
real values are predicted (2-state prediction is also made for CN using 
the average CN for each residue type as the threshold for ``exposed'' 
or ``buried'' as in \cite{PollastriETAL2002}).
The half window size $M$ is set to 9 for SS and CN predictions, and to 26 for 
RWCO. 

\subsection*{Ensemble prediction}
Since the CRN-based prediction is parametrized by the random matrices $W$ 
and $V$,
slightly different predictions are obtained for different pairs of $W$ and $V$. 
We can improve the prediction by taking the average over an ensemble of 
such different predictions. 20 CRN-based predictors were constructed using 
20 sets of different random matrices $W$ and $V$. CN and RWCO are predicted 
as uniform averages of these 20 predictions. 

For SS prediction, we employ further training. Let $s_{i}^{t,n}$ be the 
prediction results of the $n$-th predictor for 1D structure $t$ 
($H$, $E$, $C$, CN, and RWCO) of the $i$-th residue.
The second stage SS prediction is made by the following linear scheme:
\begin{equation}
  \label{eq:ss2}
  y_{i}^{ss} = \sum_{n=1}^{20}\sum_{t}\sum_{m=-3}^{3}w_{n,t,m}s_{i+m}^{t,n}
\end{equation}
where $ss = H, E, C$, and $w_{n,t,m}$ is the weight obtained from a training 
set. Finally, the feature variable for each SS class of the 
$i$-th residue is obtained by $(y_{i-1}^{ss} + 2y_{i}^{ss} + y_{i+1}^{ss})/4$. 
This last procedure was found particularly effective for improving the 
segment overlap (SOV) measure.

\subsection*{Additional input}
Another improvement is the addition of the amino acid composition of 
the target sequence to the predictor \cite{Yuan2005}:
The term $\sum_{a=1}^{20}F_af_a$ was added to Eq. \ref{eq:pred} where $F_a$
is a regression parameter, and $f_a$ is the fraction of the amino acid 
type $a$.

\subsection*{Training and test data set}
We carried out a 15-fold cross-validation test following exactly the same 
procedure and the same data set as the previous 
study \cite{KinjoANDNishikawa2005c}. In the data set, there are 680 protein 
domains, each of which represents a superfamily according to the SCOP 
database (version 1.65) \cite{SCOP}. This data set was randomly divided so 
that 630 domains were used for training and the remaining 50 domains for 
testing, and the random division was repeated 15 times. 
No pair of these domains belong to the same superfamily, and hence they are 
not expected to be homologous. Thus, the present benchmark is a very 
stringent one.

For obtaining PSSMs by running PSI-BLAST, we use the UniRef100 
(version 6.8) amino acid sequence database \cite{UniProt} containing some
3 million entries.
Also the number of iterations in PSI-BLAST homology searches was reduced 
to 3 times from 10 used in the previous study. This especially increased the 
accuracy of SS predictions. 
These results are consistent with the study of \cite{PrzybylskiANDRost2002}.

\subsection*{Numerics}
One drawback of the CRN method is the computational time required for
numerically solving the equation of state (Eq. \ref{eq:eos}).
For that purpose, instead of the Gauss-Seidel-like 
method previously used, we implemented a successive over-relaxation 
method which was found to be much more efficient.

Let $\nu$ denote the stage of iteration.
We set the initial value of the state vectors (with $\nu = 0$) as 
\begin{equation}
  \mathbf{x}_{i}^{(0)} = \tanh [\alpha V \mathbf{u}_{i}].\label{eq:init_eos}
\end{equation}
Then, for $i = 1, \cdots , L$ (in increasing order of $i$), we update 
the state vectors by
\begin{eqnarray}
  \mathbf{x}_{i}^{(2\nu+1)} \gets & \mathbf{x}_{i}^{(2\nu)} + \omega
\{\tanh [W(\mathbf{x}_{i-1}^{(2\nu+1)}\nonumber\\
& +\mathbf{x}_{i+1}^{(2\nu)}) 
+ \alpha V \mathbf{u}_{i}] - \mathbf{x}_{i}^{(2\nu)}\}.
\label{eq:feos}
\end{eqnarray}
Next, we update them in the reverse order. That is, for $i = L, \cdots , 1$ 
(in decreasing order of $i$), 
\begin{eqnarray}
  \mathbf{x}_{i}^{(2\nu+2)}  \gets & \mathbf{x}_{i}^{(2\nu+1)} + \omega 
\{\tanh [W(\mathbf{x}_{i-1}^{(2\nu+1)} \nonumber\\
& + \mathbf{x}_{i+1}^{(2\nu+2)}) 
+\alpha V \mathbf{u}_{i}] - \mathbf{x}_{i}^{(2\nu+1)}\}.
\label{eq:beos}
\end{eqnarray}
We then set $\nu \gets \nu + 1$, and iterate Eqs. (\ref{eq:feos}) and (\ref{eq:beos}) until $\{\mathbf{x}_{i}\}$ converges. The acceleration parameter of $\omega = 1.4$ was found effective. 
The convergence criterion is 
\begin{equation}
\sqrt{\sum_{i=1}^{L}||\mathbf{x}_{i}^{(2\nu+2)}-\mathbf{x}_{i}^{(2\nu+1)}||_{\mathbf{R}^{N}}^{2}/{NL}}<10^{-3}
\end{equation}
where $||\cdot||_{\mathbf{R}^{N}}$ denotes the Euclidean norm.
This criterion is much less stringent than previous study ($10^{-7}$), but this
does not affect the prediction accuracy significantly.
Convergence is typically achieved within 10 to 12 iterations for one protein.

\section*{Results and Discussion}
There are two main ingredients for the improved one-dimensional protein 
structure prediction in the present study. First is the use of large-scale 
critical random networks of 5000 dimension and 20 ensemble predictors. 
Second is the use of a large sequence database (UniRef100) for PSI-BLAST 
searches.
As demonstrated in Table~1, the CRN method achieves remarkably 
accurate predictions.
In comparison with the previous study \cite{KinjoANDNishikawa2005c} based on
2000-dimensional CRNs (10 ensemble predictors), 
the $Q_3$ and $SOV$ measures in SS predictions improved from 77.8\% and 77.3\% 
to 80.5\% and 80.0\%, respectively. Similarly, the average correlation 
coefficient improved from 0.726 to 0.746 for CN predictions, 
and from 0.601 to 0.613 for RWCO predictions. The 2-state predictions for 
CN yields, on average, $Q_2$ = 76.8\% per chain and 76.7\% per residue, and 
Matthews' correlation coefficient of 0.533.

A closer examination of the SS prediction results (Table 2) 
reveals the drastic improvement of $\beta$ strand prediction from $Q_E$ 
= 61.9\% to 69.3\% (per residue). Although the values of $Q_C$ and $Q_E^{pre}$
are slightly lower than in the previous study by 0.6--1.0\%, the accuracies of
other classes have improved by 2.5--4\%.

CRNPRED compares favorably with other secondary structure prediction methods.
The widely used PSIPRED program \cite{Jones1999,PSIPRED} which is based on conventional 
feed-forward neural networks achieves $Q_3$ of 78\%. 
A more recently developed method, Porter, \cite{PollastriANDMcLysaght2005}
which is based on bidirectional recurrent neural networks achieves $Q_3$ of 
79\%. An even more intricate method based on bidirectional segmented-memory 
recurrent neural networks \cite{ChenANDChaudhari2006} shows an accuracy 
of $Q_3$ = 73\% (this rather low accuracy may be attributed to the small size 
of training set used). However, it should be reminded that these studies are 
based on different data sets for both training and testing as well as the 
definition of 
secondary structural categories. Therefore, these comparisons may not be 
very informative, but only give a rough estimation of relative performance. 

Regarding the contact number prediction, CRNPRED, achieving $Cor$ = 0.75, 
is the most accurate method available today. The simple linear method \cite{KinjoETAL2005} with multiple 
sequence alignment derived from the HSSP database \cite{HSSP} showed a 
correlation coefficient of 0.63. A more advanced method based on support vector machines (local window-based) achieves a correlation of 0.68 per chain\cite{Yuan2005}.

It is known that the number of homologs found by the PSI-BLAST searches 
significantly affects the prediction accuracies \cite{PrzybylskiANDRost2002}. 
We have examined this effect by plotting the accuracy measures for a 
given minimum number of homologs found by PSI-BLAST (Fig. 1).
For example, we see in Fig. 1 that, for those proteins with 
more than 100 homologs, the average $Q_3$ for SS predictions is 82.2\%.  
The effect of the number of homologs significantly depends on the type of 
1D structure. For SS prediction, $Q_3$ steadily increases as the number of 
homologs increases up to 100, but it stays in the range between 82.0 and 82.4
until the minimum number of homologs reaches around 400, and then it starts to
decrease. For CN prediction, $Cor$ also increases steadily but more slowly, 
and it does not degrade when the minimum number of homologs reaches 500. 
This tendency implies that CN is more conservative than SS during protein 
evolution, which is consistent with previous observations \cite{KinjoANDNishikawa2004,BastollaETAL2005}. On the contrary, RWCO exhibits a peculiar behavior. 
The value of $Cor$ reaches its peak at the minimum number of homologs of 80 
beyond which the value rapidly decreases. This indicates that RWCO is not 
evolutionarily well conserved. It was observed that the accuracies of SS and 
CN predictions constantly increased when the dimension of CRNs was increased 
from 2000 to 5000, but such was not the case for RWCO (data not shown). 
RWCO seems to be such delicate a quantity that it is very difficult to extract 
relevant information from the amino acid sequence.

Finally, we note on practical applicability of predicted 1D 
structures. We do not believe, at present, that the construction of
a 3D structure purely from the predicted 1D structures is practical, 
if possible at all, because of the limited accuracy of the RWCO prediction.
However, SS and CN predictions are very accurate for many proteins 
so that they may already serve as valuable restraints for 3D structure 
predictions. Also, SS and CN predictions may be applied to domain 
identification often necessary for experimental determination of protein 
structures. CRNPRED has been proved useful for such a purpose \cite{MinezakiETAL2006}.
Although of the limited accuracy, predicted RWCOs still exhibit significant 
correlations with the correct values. Since RWCOs reflect the extent to which
a residue is involved in long-range contacts, predicted RWCOs may be 
useful for enumerating potentially structurally important residues. 

An interesting alternative application of the CRN framework is to regard the 
solution of the equation of state (Eq. \ref{eq:eos}) as an extended sequence 
profile. By so doing, it is straightforward to apply the solution to the 
profile-profile comparison for fold recognition \cite{TomiiANDAkiyama2004}. 
Such an application may be also pursued in the future.

\section*{Availability and Requirements}

\begin{description}
\item[Project name:] CRNPRED
\item[Project home page:] ~\\http://bioinformatics.org/crnpred/
\item[Operating system:] UNIX-like OS (including Linux and Mac OS X).
\item[Programming language:] C.
\item[Other requirements:] zsh, PSI-BLAST (blastpgp), The UniRef100 amino acid sequence database.
\item[License:] Public domain.
\item[Any restrictions to use by non-academics:] None.
\end{description}

\section*{List of Abbreviations Used}
CRN, critical random network; SS, secondary structure; CN, contact number; 
RWCO, residue-wise contact order; 1D, one-dimensional; 3D, three-dimensional.

\section*{Authors contributions}
A. R. K. designed and implemented the method, carried out benchmarks, wrote 
the first draft of the manuscript. A. R. K. and K. N. analyzed the results and 
improved the manuscript.

\section*{Acknowledgements}
We thank Yasumasa Shigemoto for helping construct the CRNPRED web interface.
This work was supported in part by the MEXT, Japan.


\newcommand{\BMCxmlcomment}[1]{}

\BMCxmlcomment{

<refgrp>

<bibl id="B1">
  <title><p>Reconstruction of protein structures from a vectorial
  representation</p></title>
  <aug>
    <au><snm>Porto</snm><fnm>M.</fnm></au>
    <au><snm>Bastolla</snm><fnm>U.</fnm></au>
    <au><snm>Roman</snm><fnm>H. E.</fnm></au>
    <au><snm>Vendruscolo</snm><fnm>M.</fnm></au>
  </aug>
  <source>Phys. Rev. Lett.</source>
  <pubdate>2004</pubdate>
  <volume>92</volume>
  <fpage>218101</fpage>
</bibl>

<bibl id="B2">
  <title><p>Recoverable one-dimensional encoding of protein three-dimensional
  structures</p></title>
  <aug>
    <au><snm>Kinjo</snm><fnm>A. R.</fnm></au>
    <au><snm>Nishikawa</snm><fnm>K.</fnm></au>
  </aug>
  <source>Bioinformatics</source>
  <pubdate>2005</pubdate>
  <volume>21</volume>
  <fpage>2167</fpage>
  <lpage>2170</lpage>
  <note>doi:10.1093/bioinformatics/bti330</note>
</bibl>

<bibl id="B3">
  <title><p>Prediction in {1D}: secondary structure, membrane helices, and
  accessibility</p></title>
  <aug>
    <au><snm>Rost</snm><fnm>B.</fnm></au>
  </aug>
  <source>Structural Bioinformatics</source>
  <publisher>Hoboken, U.S.A.: Wiley-Liss, Inc.</publisher>
  <editor>Bourne, P. E. and Weissig, H.</editor>
  <section><title><p>28</p></title></section>
  <pubdate>2003</pubdate>
  <fpage>559</fpage>
  <lpage>587</lpage>
</bibl>

<bibl id="B4">
  <title><p>Protein secondary structure prediction based on position-specific
  scoring matrices</p></title>
  <aug>
    <au><snm>Jones</snm><fnm>D. T.</fnm></au>
  </aug>
  <source>J. Mol. Biol.</source>
  <pubdate>1999</pubdate>
  <volume>292</volume>
  <fpage>195</fpage>
  <lpage>202</lpage>
</bibl>

<bibl id="B5">
  <title><p>Porter: a new, accurate server for protein secondary structure
  prediction</p></title>
  <aug>
    <au><snm>Pollastri</snm><fnm>G.</fnm></au>
    <au><snm>{McLysaght}</snm><fnm>A.</fnm></au>
  </aug>
  <source>Bioinformatics</source>
  <pubdate>2005</pubdate>
  <volume>21</volume>
  <fpage>1719</fpage>
  <lpage>-1720</lpage>
</bibl>

<bibl id="B6">
  <title><p>Prediction of the surface-interior diagram of globular proteins by
  an empirical method</p></title>
  <aug>
    <au><snm>Nishikawa</snm><fnm>K.</fnm></au>
    <au><snm>Ooi</snm><fnm>T.</fnm></au>
  </aug>
  <source>Int. J. Peptide Protein Res.</source>
  <pubdate>1980</pubdate>
  <volume>16</volume>
  <fpage>19</fpage>
  <lpage>32</lpage>
</bibl>

<bibl id="B7">
  <title><p>Radial locations of amino acid residues in a globular protein:
  Correlation with the sequence</p></title>
  <aug>
    <au><snm>Nishikawa</snm><fnm>K.</fnm></au>
    <au><snm>Ooi</snm><fnm>T.</fnm></au>
  </aug>
  <source>J. Biochem.</source>
  <pubdate>1986</pubdate>
  <volume>100</volume>
  <fpage>1043</fpage>
  <lpage>1047</lpage>
</bibl>

<bibl id="B8">
  <title><p>Predicting absolute contact numbers of native protein structure
  from amino acid sequence</p></title>
  <aug>
    <au><snm>Kinjo</snm><fnm>A. R.</fnm></au>
    <au><snm>Horimoto</snm><fnm>K.</fnm></au>
    <au><snm>Nishikawa</snm><fnm>K.</fnm></au>
  </aug>
  <source>Proteins</source>
  <pubdate>2005</pubdate>
  <volume>58</volume>
  <fpage>158</fpage>
  <lpage>165</lpage>
  <note>doi:10.1002/prot.20300</note>
</bibl>

<bibl id="B9">
  <title><p>Better prediction of protein contact number using a support vector
  regression analysis of amino acid sequence</p></title>
  <aug>
    <au><snm>Yuan</snm><fnm>Z.</fnm></au>
  </aug>
  <source>BMC Bioinformatics</source>
  <pubdate>2005</pubdate>
  <volume>6</volume>
  <fpage>248</fpage>
</bibl>

<bibl id="B10">
  <title><p>Predicting secondary structures, contact numbers, and residue-wise
  contact orders of native protein structure from amino acid sequence using
  critical random networks</p></title>
  <aug>
    <au><snm>Kinjo</snm><fnm>A. R.</fnm></au>
    <au><snm>Nishikawa</snm><fnm>K.</fnm></au>
  </aug>
  <source>BIOPHYSICS</source>
  <pubdate>2005</pubdate>
  <volume>1</volume>
  <fpage>67</fpage>
  <lpage>74</lpage>
  <note>doi:10.2142/biophysics.1.67</note>
</bibl>

<bibl id="B11">
  <title><p>Dictionary of Protein Secondary Structure: Pattern recognition of
  hydrogen bonded and geometrical features</p></title>
  <aug>
    <au><snm>Kabsch</snm><fnm>W.</fnm></au>
    <au><snm>Sander</snm><fnm>C.</fnm></au>
  </aug>
  <source>Biopolymers</source>
  <pubdate>1983</pubdate>
  <volume>22</volume>
  <fpage>2577</fpage>
  <lpage>2637</lpage>
</bibl>

<bibl id="B12">
  <title><p>Protein secondary structure: entropy, correlations and
  prediction</p></title>
  <aug>
    <au><snm>Crooks</snm><fnm>G. E.</fnm></au>
    <au><snm>Brenner</snm><fnm>S. E.</fnm></au>
  </aug>
  <source>Bioinformatics</source>
  <pubdate>2004</pubdate>
  <volume>20</volume>
  <fpage>1603</fpage>
  <lpage>1611</lpage>
</bibl>

<bibl id="B13">
  <title><p>Exploiting the past and the future in protein secondary structure
  prediction</p></title>
  <aug>
    <au><snm>Baldi</snm><fnm>P.</fnm></au>
    <au><snm>Brunak</snm><fnm>S.</fnm></au>
    <au><snm>Frasconi</snm><fnm>P.</fnm></au>
    <au><snm>Soda</snm><fnm>G.</fnm></au>
    <au><snm>Pollastri</snm><fnm>G.</fnm></au>
  </aug>
  <source>Bioinformatics</source>
  <pubdate>1999</pubdate>
  <volume>15</volume>
  <fpage>937</fpage>
  <lpage>946</lpage>
</bibl>

<bibl id="B14">
  <title><p>Bidirectional segmented-memory recurrent neural network for protein
  secondary structure prediction</p></title>
  <aug>
    <au><snm>Chen</snm><fnm>J.</fnm></au>
    <au><snm>Chaudhari</snm><fnm>N. S.</fnm></au>
  </aug>
  <source>Soft Computing</source>
  <pubdate>2006</pubdate>
  <volume>10</volume>
  <fpage>315</fpage>
  <lpage>324</lpage>
</bibl>

<bibl id="B15">
  <title><p>Gapped Blast and {PSI}-Blast: A new generation of protein database
  search programs</p></title>
  <aug>
    <au><snm>Altschul</snm><fnm>S. F.</fnm></au>
    <au><snm>Madden</snm><fnm>T. L.</fnm></au>
    <au><snm>Schaffer</snm><fnm>A. A.</fnm></au>
    <au><snm>Zhang</snm><fnm>J.</fnm></au>
    <au><snm>Zhang</snm><fnm>Z.</fnm></au>
    <au><snm>Miller</snm><fnm>W.</fnm></au>
    <au><snm>Lipman</snm><fnm>D. L.</fnm></au>
  </aug>
  <source>Nucleic Acids Res.</source>
  <pubdate>1997</pubdate>
  <volume>25</volume>
  <fpage>3389</fpage>
  <lpage>3402</lpage>
</bibl>

<bibl id="B16">
  <title><p>Prediction of coordination number and relative solvent
  accessibility in proteins</p></title>
  <aug>
    <au><snm>Pollastri</snm><fnm>G.</fnm></au>
    <au><snm>Baldi</snm><fnm>P.</fnm></au>
    <au><snm>Fariselli</snm><fnm>P.</fnm></au>
    <au><snm>Casadio</snm><fnm>R.</fnm></au>
  </aug>
  <source>Proteins</source>
  <pubdate>2002</pubdate>
  <volume>47</volume>
  <fpage>142</fpage>
  <lpage>153</lpage>
</bibl>

<bibl id="B17">
  <title><p>{SCOP}: A structural classification of proteins database for the
  investigation of sequences and structures</p></title>
  <aug>
    <au><snm>Murzin</snm><fnm>A. G.</fnm></au>
    <au><snm>Brenner</snm><fnm>S. E.</fnm></au>
    <au><snm>Hubbard</snm><fnm>T.</fnm></au>
    <au><snm>Chothia</snm><fnm>C.</fnm></au>
  </aug>
  <source>J. Mol. Biol.</source>
  <pubdate>1995</pubdate>
  <volume>247</volume>
  <fpage>536</fpage>
  <lpage>540</lpage>
</bibl>

<bibl id="B18">
  <title><p>The universal protein resource ({UniProt})</p></title>
  <aug>
    <au><snm>Bairoch</snm><fnm>A.</fnm></au>
    <au><snm>Apweiler</snm><fnm>R.</fnm></au>
    <au><snm>Wu</snm><fnm>C. H.</fnm></au>
    <au><snm>Barker</snm><fnm>W. C.</fnm></au>
    <au><snm>Boeckmann</snm><fnm>B.</fnm></au>
    <au><snm>Ferro</snm><fnm>S.</fnm></au>
    <au><snm>Gasteiger</snm><fnm>E.</fnm></au>
    <au><snm>Huang</snm><fnm>H.</fnm></au>
    <au><snm>Lopez</snm><fnm>R.</fnm></au>
    <au><snm>Magrane</snm><fnm>M.</fnm></au>
    <au><snm>Martin</snm><fnm>M. J.</fnm></au>
    <au><snm>Natale</snm><fnm>D.A.</fnm></au>
    <au><snm>{O'Donovan}</snm><fnm>C.</fnm></au>
    <au><snm>Redaschi</snm><fnm>N.</fnm></au>
    <au><snm>Yeh</snm><fnm>L. S.</fnm></au>
  </aug>
  <source>Nucleic Acids Res.</source>
  <pubdate>2005</pubdate>
  <volume>33</volume>
  <fpage>D154</fpage>
  <lpage>D159</lpage>
</bibl>

<bibl id="B19">
  <title><p>Alignments grow, secondary structure prediction
  improves</p></title>
  <aug>
    <au><snm>Przybylski</snm><fnm>D.</fnm></au>
    <au><snm>Rost</snm><fnm>B.</fnm></au>
  </aug>
  <source>Proteins</source>
  <pubdate>2002</pubdate>
  <volume>46</volume>
  <fpage>197</fpage>
  <lpage>205</lpage>
</bibl>

<bibl id="B20">
  <title><p>The PSIPRED protein structure prediction server</p></title>
  <aug>
    <au><snm>McGuffin</snm><fnm>L. J.</fnm></au>
    <au><snm>Bryson</snm><fnm>K.</fnm></au>
    <au><snm>Jones</snm><fnm>D. T.</fnm></au>
  </aug>
  <source>Bioinformatics</source>
  <pubdate>2000</pubdate>
  <volume>16</volume>
  <fpage>404</fpage>
  <lpage>405</lpage>
</bibl>

<bibl id="B21">
  <title><p>Database of homology-derived protein structures</p></title>
  <aug>
    <au><snm>Sander</snm><fnm>C.</fnm></au>
    <au><snm>Schneider</snm><fnm>R.</fnm></au>
  </aug>
  <source>Proteins</source>
  <pubdate>1991</pubdate>
  <volume>9</volume>
  <fpage>56</fpage>
  <lpage>68</lpage>
</bibl>

<bibl id="B22">
  <title><p>Eigenvalue analysis of amino acid substitution matrices reveals a
  sharp transition of the mode of sequence conservation in proteins</p></title>
  <aug>
    <au><snm>Kinjo</snm><fnm>A. R.</fnm></au>
    <au><snm>Nishikawa</snm><fnm>K.</fnm></au>
  </aug>
  <source>Bioinformatics</source>
  <pubdate>2004</pubdate>
  <volume>20</volume>
  <fpage>2504</fpage>
  <lpage>2508</lpage>
</bibl>

<bibl id="B23">
  <title><p>Principal eigenvector of contact matrices and hydrophobicity
  profiles in proteins</p></title>
  <aug>
    <au><snm>Bastolla</snm><fnm>U.</fnm></au>
    <au><snm>Porto</snm><fnm>M.</fnm></au>
    <au><snm>Roman</snm><fnm>H. E.</fnm></au>
    <au><snm>Vendruscolo</snm><fnm>M.</fnm></au>
  </aug>
  <source>Proteins</source>
  <pubdate>2005</pubdate>
  <volume>58</volume>
  <fpage>22</fpage>
  <lpage>30</lpage>
</bibl>

<bibl id="B24">
  <title><p>Human transcription factors contain a high fraction of
  intrinsically disordered regions essential for transcriptional
  regulation</p></title>
  <aug>
    <au><snm>Minezaki</snm><fnm>Y.</fnm></au>
    <au><snm>Homma</snm><fnm>K.</fnm></au>
    <au><snm>Kinjo</snm><fnm>A. R.</fnm></au>
    <au><snm>Nishikawa</snm><fnm>K.</fnm></au>
  </aug>
  <source>J. Mol. Biol.</source>
  <pubdate>2006</pubdate>
  <inpress />
</bibl>

<bibl id="B25">
  <title><p>{FORTE}: a profile-profile comparison tool for protein fold
  recognition</p></title>
  <aug>
    <au><snm>Tomii</snm><fnm>K.</fnm></au>
    <au><snm>Akiyama</snm><fnm>Y.</fnm></au>
  </aug>
  <source>Bioinformatics</source>
  <pubdate>2004</pubdate>
  <volume>20</volume>
  <fpage>594</fpage>
  <lpage>595</lpage>
</bibl>

<bibl id="B26">
  <title><p>A modified definition of Sov, a segment-based measure for protein
  secondary structure prediction assessment</p></title>
  <aug>
    <au><snm>Zemla</snm><fnm>A</fnm></au>
    <au><snm>Venclovas</snm><fnm>C.</fnm></au>
    <au><snm>Fidelis</snm><fnm>K.</fnm></au>
    <au><snm>Rost</snm><fnm>B.</fnm></au>
  </aug>
  <source>Proteins</source>
  <pubdate>1999</pubdate>
  <volume>34</volume>
  <fpage>220</fpage>
  <lpage>223</lpage>
</bibl>

</refgrp>
} 

\newpage
\section*{Figures}
  \subsection*{Figure 1}
Average accuracy measure for given minimum number of homologs found by PSI-BLAST. From top to bottom: $Q_3$ of secondary structure predictions, $Cor$ of contact number predictions, and $Cor$ of residue-wise contact number predictions.

\includegraphics[width=6cm]{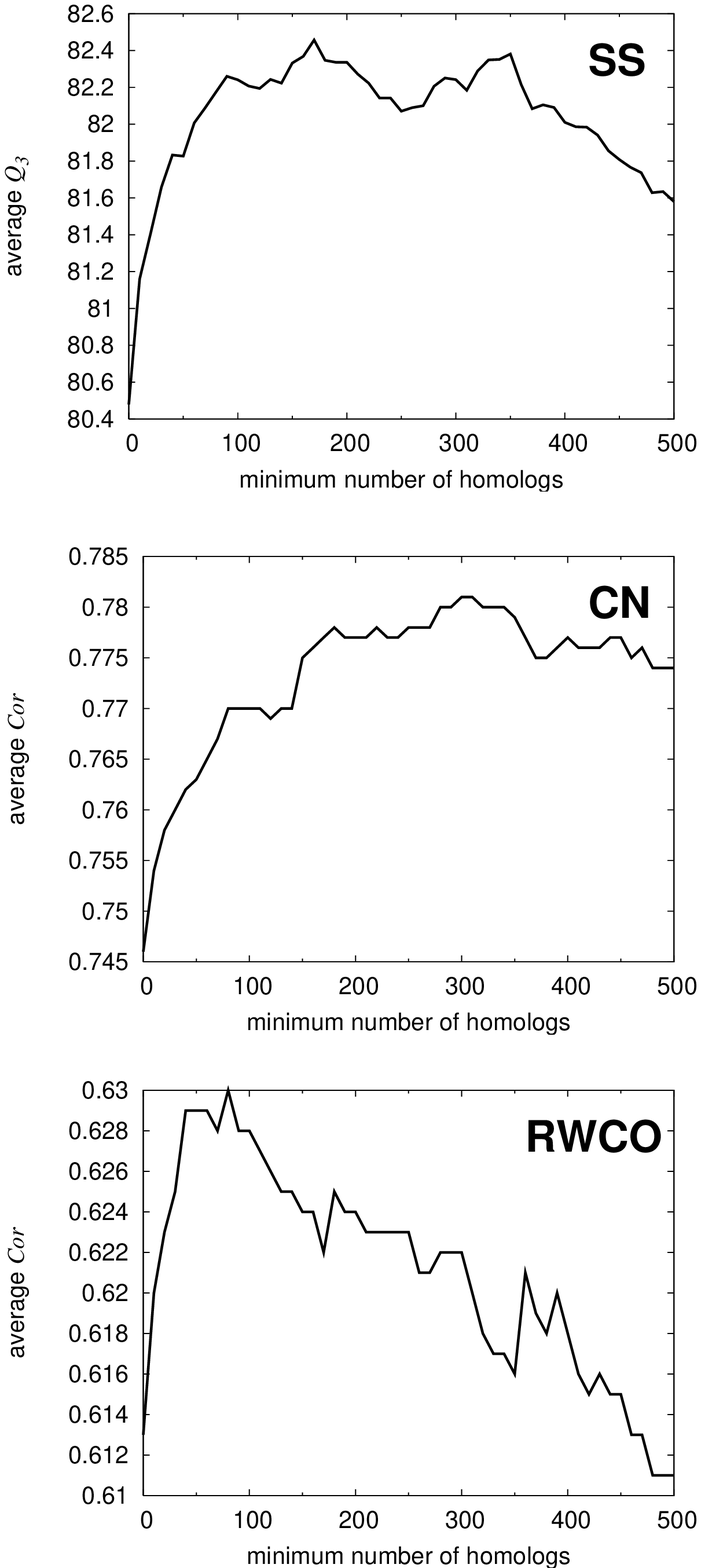}

\newpage

\section*{Tables}
  \subsection*{Table 1 - Summary of average prediction accuracies per chain (median in parentheses).}
    \par
    \mbox{
\begin{tabular}{lll}\hline
SS & $Q_3$= 80.5\% (81.6) & $SOV$= 80.0\% (81.1)\\
CN & $Cor$= 0.746 (0.768) & $DevA$= 0.686 (0.670) \\
RWCO & $Cor$= 0.613 (0.646) & $DevA$= 0.877 (0.812)\\\hline
  \end{tabular}
      }\\

SS, Secondary structure prediction: $Q_3$ is the percentage of correct prediction.; $SOV$ is the segment overlap measure~\cite{SOV99}.\\
CN, Contact number prediction: $Cor$ is the Pearson's correlation coefficient between the predicted and native CNs; $DevA$ is the RMS error normalized by the standard deviation of the native CN \cite{KinjoETAL2005}.\\
RWCO, Residue-wise contact order prediction: $Cor$ and $DevA$ are defined as for
CN but calculated with predicted and native RWCOs.

\subsection*{Table 2: Summary of per-residue accuracies for SS predictions.}
\par
\mbox{
  \begin{tabular}[tbh]{lrrr}\hline
measure    & $H$ & $E$ & $C$ \\\hline
$Q_s$      & 82.7 & 69.3 & 84.0 \\
$Q_s^{pre}$ & 84.4 & 78.9 & 78.3\\
$MC$       &  0.754 & 0.674 & 0.645 \\\hline
  \end{tabular}
}\\

$Q_s$: The number of correctly predicted residues of the SS class $s = H, E, C$
 divided by the number of residues in the class in native structures.\\
$Q_s^{pre}$: The number of correctly predicted residues of the SS class $s = H, E, C$
 divided by the number of residues predicted as the corresponding class.\\
$MC$: Matthews' correlation coefficient.
\end{document}